\documentclass[11pt]{article}

\usepackage[english]{babel}
\usepackage{setspace}
\usepackage[margin=2.5cm]{geometry}
\usepackage{authblk}
\usepackage{amsfonts}
\usepackage{amsmath}
\usepackage{graphicx}
\usepackage{url}
\usepackage{subfigure}
\usepackage{algorithm}
\usepackage{algpseudocode}
\usepackage[square,numbers]{natbib}
\usepackage{calc}

\newcommand{\beq}{\begin{equation}}
\newcommand{\eeq}{\end{equation}}
\newcommand{\Fig}[1]{Fig-\ref{#1}}
\newcommand{\fig}[1]{fig-\ref{#1}}

\title{Realistic voxel sizes and reduced signal variation in Monte-Carlo simulation for diffusion MR data synthesis}

\author{Matt G Hall}
\author{Gemma Nedjati-Gilani}
\author{Daniel C Alexander}
\affil{\small{Center for Medical Image Computing\\
University College London\\
Gower Street\\
London WC1E 6BT\\
United Kingdom}}

\begin{document}

\urlstyle{sf}

\maketitle

\onehalfspacing

\begin{abstract}
To synthesize diffusion MR measurements from Monte-Carlo simulation using tissue models with sizes comparable to those of scan voxels. Larger regions enable restricting structures to be modeled in greater detail and improve accuracy and precision in synthesized diffusion-weighted measurements. 

We employ a localized intersection checking algorithm during substrate construction and dynamical simulation. Although common during dynamics simulation, a dynamically constructed intersection map is also applied here during substrate construction, facilitating construction of much larger substrates than would be possible with a naive ``brute-force'' approach. We investigate the approach's performance using a packed cylinder model of white matter, investigating optimal execution time for simulations, convergence of synthesized signals and variance in diffusion-weighted measurements over a wide range of acquisition parameters. The scheme is demonstrated with cylinder-based substrates but is also readily applicable to other geometric primitives, such as spheres or triangles.

The algorithm enables models with far larger substrates to be run with no additional computational cost. The improved sampling reduces bias and variance in synthetic measurements. 

        The new method improves accuracy, precision, and reproducibility of synthetic measurements in Monte-Carlo simulation-based data synthesis. The larger substrates it makes possible are better able to capture the complexity of the tissue we are modeling, leading to reduced bias and variance in synthesised data, compared to existing implementation of MC simulations.
\end{abstract}


\section*{Introduction}
Diffusion-weighted MRI (DW-MRI) is an imaging modality that is sensitive to local incoherent motion of spins in a sample. DW-MRI works by attenuating the observed spin-echo by an amount related to the ensemble mean squared-displacement - broadly speaking, the further spins diffuse, the more the signal will be attenuated. The presence of barriers inhibits spins' motion and thus measurements of diffusion contain information about restricting structure in the local environment. Exploiting this information to provide in vivo estimates of tissue microstructure is currently a major avenue of diffusion imaging research.

One particular challenge to the development of more accurate reconstruction techniques is the need for detailed ground truth information against which the analysis can be validated. One approach to this is Monte-Carlo simulation. This allows synthetic DW-MRI measurements to be generated by simulating trajectories of spins executing random walks in an environment that may be specified in detail as a known model of tissue environment. The tissue environment can be comprised of geometric primitives, such as cylinders and spheres, or derived from biological images and represented as a triangle mesh \cite{LauraMICCAI2010}. Simulation provides a cheap and accessible method of investigating computational phantoms capable of capturing a great deal of the complexity of biological tissue in a controlled way. 

Over the last few years Monte-Carlo simulation has become increasingly widely-used. For example to investigate the effect of tissue swelling on DW-MRI signals \cite{hallAlexander09}, the reconstruction of crossing fibers \cite{alonso}, the effects of ``beading'' in cylinders \cite{budde}, the effect of fiber undulation on diffusion-weighted data \cite{nilsson}, the effect of cylinder wall-thickness \cite{BallFranks}, the viability of estimating cylinder radius distribution parameters from DW-MRI data \cite{activeAx}, and the validity of novel theoretical models of diffusion \cite{NovikovNatPhys}.

In each of these studies, parameter estimates from synthetic diffusion-weighted MR data are compared to the ground truth used in the simulation where the micro-structural parameters are known and defined a priori. This enables the bias and variance in parameter estimates to be assessed and quantified - a crucial step in evaluating a new technique. In each case, diffusion is simulated in a chosen tissue model. We will refer to this tissue model as a ``substrate''.

Most Monte-Carlo simulations published so far have suffered from the shortcoming that the substrate used in the simulation is considerably smaller than a typical scan voxel. This is important when considering complex and realistic tissue environments because a small substrate cannot capture the full complexity of the tissue. Here we propose a more efficient implementation of the simulation, using optimized intersection checking, that removes the limit on the complexity of the substrate and allows us to run MC simulations on voxel-sized substrates. Optimised intersection checking is common is Monte-Carlo simulation during the main biophysical dynamics of the simulation. Here we outline an approach that also employs it during substrate construction, allowing considerable improvements in performance in this phase as well.

The remainder of this article is constructed as follows. The Methods section describes the substrate, substrate construction, and intersection-checking algorithm. The experiments section then reports numerical results that demonstrate the feasibility of running simulations on voxel-sized substrates. Specifically, we quantify computation times and signal bias and variance as a function of substrate size to demonstrate the added value of the new implementation. We conclude with some recommendations of usage of the new implementation and a discussion of future work.

\section*{Methods}
\label{methods}
To investigate optimized intersection checking and its impact on diffusion MR data synthesis we consider a standard model of white matter: randomly packed, non-overlapping cylinders with gamma-distributed radii. This has been used in several previous studies e.g. \cite{hallAlexander09}, \cite{activeAx}.

Simulations first construct the substrate by arranging a set of cylinders in a region of a chosen size, as in algorithm 1. Substrates are constructed at run time by specifying the cylinder distribution parameters, size of spatial region and number of cylinders. Once constructed, the simulation runs random walks within the substrate in which each spin makes a fixed length step in a random direction. Any intersection with a cylinder boundary results in an elastic reflection in which the total length of the step is conserved.

\begin{algorithm}
	\begin{algorithmic}[1]
		\State Draw a sample $R$ of $M$ radii from gamma distribution $\Gamma\left(k,\theta\right)$
		\State Sort into descending order

		\ForAll {$r \in R$}
			\State $count \gets 0$
			\While {$reject = \texttt{True}$}
				\State Generate trial location $u \gets \left(u_x,u_y\right)$
				\State $reject \gets \texttt{False}$
				\If {Cylinder intersects any other cylinders on the substrate}
					\State $count \gets count+1$
					\State $reject \gets \texttt{True}$
					\State \textbf{break}
				\EndIf

				\If {$count \geq$ MAXCOUNT}
					\State $reject \gets \texttt{False}$
				\EndIf
			\EndWhile

			\If {$count \geq$ MAXCOUNT}
				\State Generate error message; abort
			\EndIf

			\State Place cylinder on substrate

		\EndFor
	\end{algorithmic}
\caption{Simplified algorithm for cylinder substrate construction ignoring steps that ensure periodicity at substrate boundary; see \cite{hallAlexander09} for full details.}
\label{alg:substrate}
\end{algorithm}

It is necessary to check intersections during substrate construction, to ensure there are no intersections between separate cylinders, and during the simulation dynamics, to detect intersections between spins and cylinders and thus enforce elastic reflections. The implementations in \cite{hallAlexander09}\cite{ElsSim}\cite{BallFranks} check each cylinder so each spin update has $O(M)$ checks, where $M$ is the number of cylinders. We refer to this as a ``brute-force'' approach. 

Intersection checking is a problem frequently encountered in computer graphics, where it is important in applications such as ray tracing and visibility determination. To increase algorithmic performance in these applications, a successful approach has been to subdivide the space into smaller regions mapped to the objects they contain. Intersection checking is then performed only for objects in relevant sub-regions. This greatly reduces the number of intersection checks necessary at one time by eliminating very distant objects that cannot possibly interact. The approach we adopt here is to employ a uniform spatial subdivision which divides the substrate into a uniform grid. The number of explicit intersection checks necessary is reduced by first assembling a list of candidate sub-regions. By doing this, the number of checks thus becomes independent of $M$, enabling simulations on much larger substrates limited only by available computer memory. This approach is described in detail in the next section.

\subsection*{Uniform spatial subdivision}
\label{spacopt}
Uniform spatial subdivision is an algorithm due to \cite{fujimoto1986}\cite{Amanatides1987} and \cite{ClearyWyvill1988}. It works by dividing the substrate into a regular grid of sub-voxels of a chosen size. Each sub-voxel is associated with the set of objects that intersect it. A new object or step on the substrate can be readily mapped to a set of sub-voxels and intersections checked only against objects intersecting these regions, constructing a map of sub-voxels to objects.

Constructing this map is a non-trivial procedure but by starting with an empty map and incrementally adding new objects to it as the substrate is constructed enables us to make use of the uniform spatial subdivision while building the substrate and benefit from the associated performance gain.

\begin{figure}[h!]
	\begin{center}
		\includegraphics[width=0.2\textwidth]{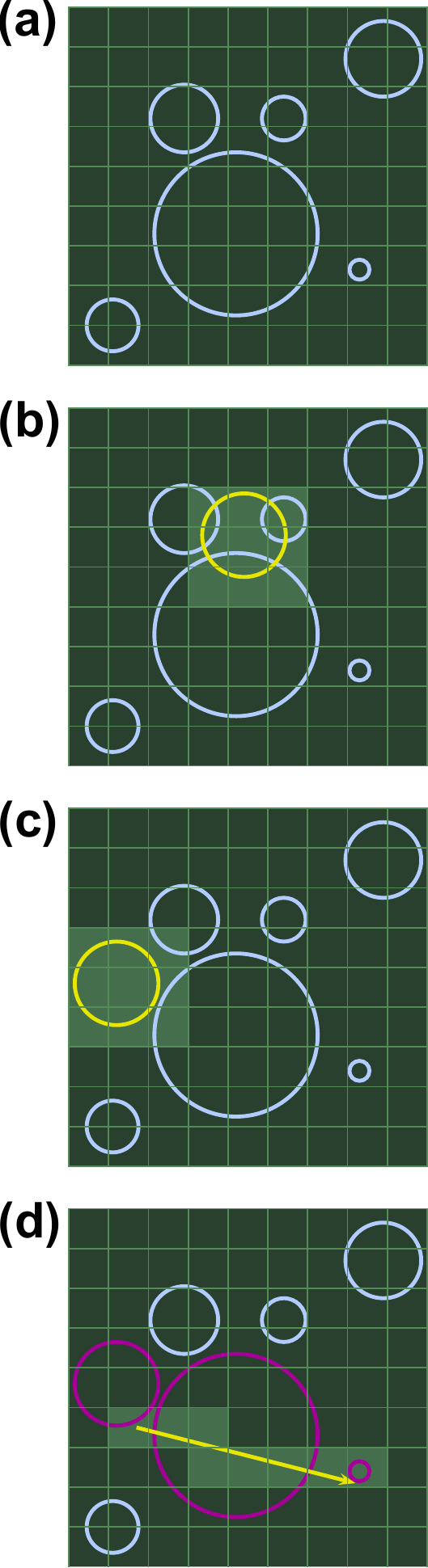}
		\caption{Uniform spatial subdivision dynamic mapping for substrate construction. The spatial region of the substrate is divided into a regular grid, and a map of sub-voxels (i.e. grid elements) to cylinders placed on the substrate constructed. When a new object is added to the substrate, it is tested for intersection with the objects already present. (a) shows the configurations of objects on the substrate. (b) The new object is tested for intersection against other objects intersecting the same sub-voxels. An explicit check is performed only against objects intersecting the light-green subvoxels. In this case intersections are detected and the location is rejected. (c) The same object is now checked at a new location against other objects intersecting the new location - this time no intersections are detected and the location is accepted. Each step in a spin's trajectory intersects one or more sub-voxels (d), and these are used to assemble a list of objects to check intersection against.}
		\label{dynmap_cyl}
	\end{center}
\end{figure}

\Fig{dynmap_cyl} illustrates the process by which a new object is added to the substrate. \Fig{dynmap_cyl}(a) shows the configuration of cylinders at a typical step in substrate construction and the sub-voxel grid. \Fig{dynmap_cyl}(b) shows an attempt to place a new cylinder on the substrate - sub-voxels intersected are highlighted. This requires three checks rather than the six that would be required by a brute-force approach as three existing cylinders are referenced from the grid elements intersecting the candidate. In this case intersections are detected and the candidate location is rejected. \Fig{dynmap_cyl}(c) shows the same cylinder at a new location. Once again intersection checks are performed only against other objects intersecting common sub-voxels. This time no intersections are detected, the location is accepted, and the new object is added to the map. 

Once all objects have been placed on the substrate the map is complete and may be used during the simulation of spin dynamics. Once again spatial subdivision reduces the number of explicit intersection checks performed. Each spin position update is a vector which intersects a set of sub-voxels on the map. An explicit check is only necessary against objects intersecting that set of sub-voxels. It is important that the set of intersecting sub-voxels be minimal and exclude as many non-intersecting voxels as possible. To construct this set we make use of Breshenham's algorithm. This traces a step from its starting point through its intersections with sub-voxel boundaries to the end of the line. We employ a straightforward 3D generalization of the algorithm described in \cite{slater} (See chapter 17, section 3 for details). 

This procedure is illustrated in \fig{dynmap_cyl}(d), which shows the set of sub-voxels intersected by a step in a spin's trajectory as obtained from Breshenham's algorithm. Intersection is explicitly checked only against objects which intersect this set of sub-voxels (shown in purple) and a set of intersections is obtained.

\begin{figure}[h!]
	\begin{center}
		\subfigure[]{
			\label{100cyls}
			\includegraphics[width=\textwidth*\real{0.106896552}*\real{0.45}]{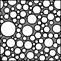}
		}
		\\
		\subfigure[]{
			\label{10000cyls}
			\includegraphics[width=0.45\textwidth]{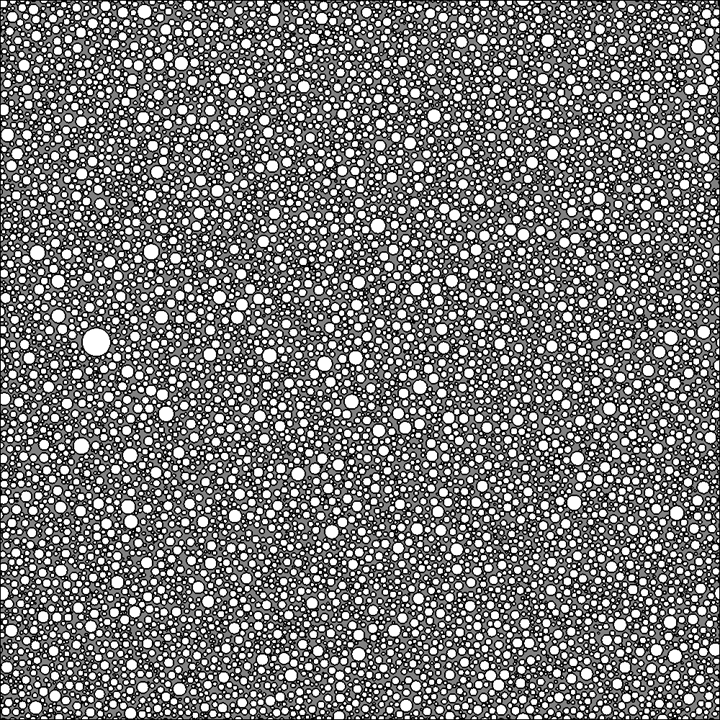}
		}
		\caption{Cross section of substrates of different sizes shown at same scale. (a) is typical of substrate sizes in previous work ($M=100$) whereas (b) contain one hundred times more objects ($M=10,000$).}
	\end{center}
\end{figure}

Use of the spatial subdivision algorithm allows simulations on substrates with far larger numbers of objects than would otherwise be possible. We illustrate this difference in fig-2. \Fig{100cyls} shows a substrate containing 100 cylinders with gamma-distributed radii - typical of previous simulation work such as \cite{hallAlexander09}. \Fig{10000cyls} shows a 10,000 cylinder substrate with cylinders drawn from the same distribution at the same spatial scale. The largest substrates used in this investigation contain 1,000,000 cylinders. These substrates are difficult to visualize directly, but a 1,000,000 cylinder substrate with cylinder radii drawn from the same distribution would occupy a region approximately ten times larger than the substrate in \fig{10000cyls} (i.e. it has 100 times the area).

\section*{Experiments}
\label{expts}
This section explores the impact of uniform spatial subdivision on substrates and synthetic data. We show how the new implementation can be optimized and explore the effect of substrate size on simulation run-time, radius distribution parameter estimation, and synthesized diffusion-weighted measurements. 

We first consider the uniform spatial subdivision algorithm itself, by investigating how to choose the size of the subdivisions in order to minimize simulation run time. We then consider the convergence properties of increasingly large substrates to the target distribution of cylinder sizes. We examine how larger substrates affect the choice of the number of spins and updates in a simulation in order to minimize bias and variation in the signals. Finally, we examine how signal variance changes over a wide range of acquisition parameters, giving an overview of the effect of spatial subdivision and larger substrates on the entire simulation process.

Throughout these experiments, substrates consist of $M$ randomly-packed non-overlapping parallel cylinders, packed into a region of size $L\times L\times L$ with radii drawn from a gamma distribution, $\Gamma(k, \theta)$. Gamma-distributed random numbers are generated using Best's Algorithm XG \cite{Best1978}, a rejection sampler method which exhibits uniform performance across all parameter values \cite{devroye1986}.

\subsection*{Choice of sub-voxel grid size}
\label{calibrate:subvox}
One key issue regarding the uniform spatial subdivision procedure is how to choose the size of the sub-voxels. In this section we examine the performance of Monte-Carlo simulations with setups identical but for the size of the sub-voxels. 
We perform the experiment for eight combinations of cylinder radius distribution and cylinder volume fraction $V_I$ to ensure that we test a wide range of substrate parameters. Cylinder radii are drawn from gamma distributions $\Gamma(k, \theta)$ with the following shape parameters $k$ and scale parameters $\theta$:
\begin{enumerate}
\item $k=5.92$, $\theta=1.06\times10^{-7}$m, (mean radius 0.63$\times10^{-6}$m, standard deviation 0.26$\times10^{-6}$m),
\item $k=6.88$, $\theta=1.70\times10^{-7}$m, (mean radius 1.17$\times10^{-6}$m, standard deviation of 0.45$\times10^{-6}$ m),
\item $k=11.0$, $\theta=2.50\times10^{-7}$m, (mean radius 2.75$\times10^{-6}$m, standard deviation of 0.83$\times10^{-6}$ m),
\item $k=15.7$, $\theta=3.40\times10^{-7}$m, (mean radius 5.34$\times10^{-6}$m, standard deviation of 1.35$\times10^{-6}$ m).
\end{enumerate}
In the first set of experiments, we maximize the cylinder volume fraction by choosing the substrate size $L$ so that 1$\times$10$^6$ randomly packed cylinders just fit without abutting. Substrate sizes for the four distinct radius distributions are $L = 1.46\times10^{-3}, 2.69\times10^{-3}, 6.27\times10^{-3}$, and $1.21\times10^{-2}$m respectively, giving cylinder volume fractions of 0.68, 0.68, 0.67 and 0.65. In the second set of experiments, we reduce the packing density to approximately 0.4 by keeping $L$ constant and reducing the number of cylinders to 6$\times10^5$. This results in cylinder volume fractions of 0.41, 0.41, 0.40 and 0.39 respectively.  

We perform experiments using 9 different sub-voxel sizes related to the mean diameter $\bar{D}$ and mode diameter $Mo(D)$ of the cylinder distributions: $\frac{\bar{D}}{4}, Mo(\frac{D}{2}), \frac{\bar{D}}{2}, Mo(D), \bar{D}, 2\bar{D}, 5\bar{D}, 10\bar{D}, 20\bar{D}$. Simulations for each sub-voxel size are performed using 100000 spins and 1000 time steps. Simulations are run 10 times, each using a different random seed to initialize the cylinder placement and dynamics. 
All simulations run on identical Linux-based nodes with Intel Xeon-E3 1240v2 processors and 16Gbytes of RAM. All code is in Java, running on JVM 1.6.0\_16, using Java Hotspot 64-bit server VM.

\begin{figure}[h!]
  \begin{center}
    \subfigure[High cylinder volume fraction]{
      \label{subvoxgridsize:v_60}
      \includegraphics[width=0.85\textwidth]{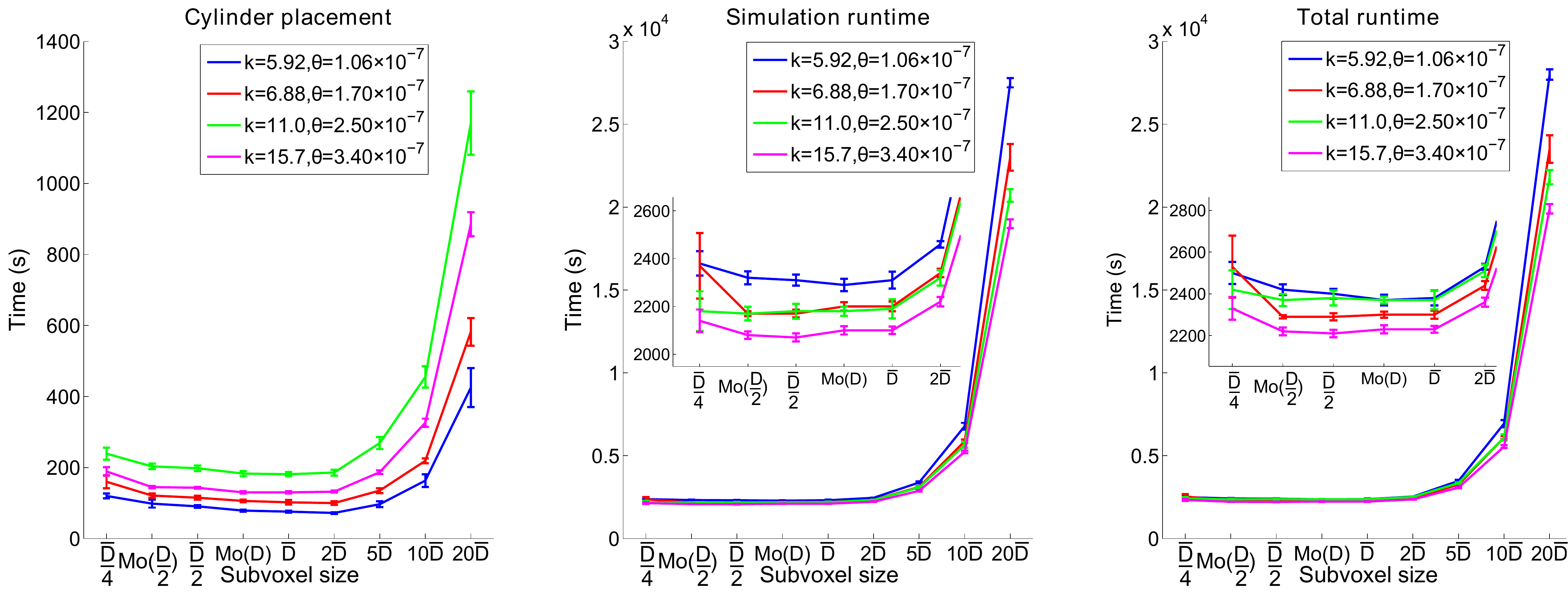}
    }
    \subfigure[Low cylinder volume fraction]{
      \label{subvoxgridsize:v_40}
      \includegraphics[width=0.85\textwidth]{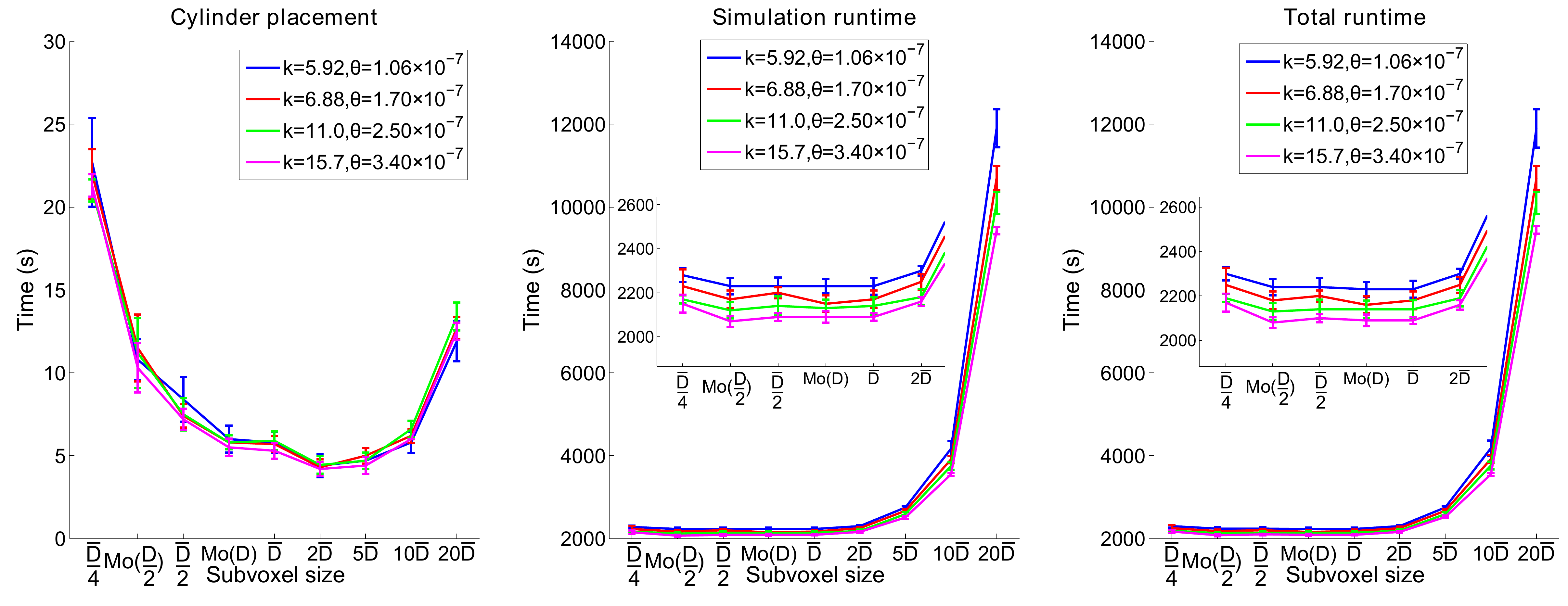}
    }
    \caption{Simulation run times as a function of sub-voxel grid size for high and low packing densities. We observe broad minima in the overall run-times at sub-voxel sizes consistent with the mode cylinder radius. This is consistent for both high and low cylinder volume fractions and a range of physically plausible cylinder radius distributions.}
    \label{subvoxgridsize}
	\end{center}
\end{figure}

\Fig{subvoxgridsize:v_60} shows the effect of sub-voxel grid size on the time taken to (left column) place the cylinders, (center column) run the simulation dynamics and (right column) total run time for all cylinder radius distributions when $V_I\approx0.70$. Minimum cylinder placement time corresponds to a sub-voxel size of $2\bar{D}$ for the two smallest cylinder distributions and $\bar{D}$ for the larger cylinders. Minimum simulation time corresponds to a sub-voxel size of $Mo(\frac{D}{2})$ for all cylinder distributions except the smallest, which is minimized for a sub-voxel size of $Mo(D)$. As the simulation time dominates total run time, the sub-voxel size that minimizes overall run time is the same as for the minimum simulation time; however the minimum is very flat over a wide range of sub-voxel sizes, from $\frac{\bar{D}}{4}$ to $2\bar{D}$. Most importantly, the shape of the curves is nearly identical for the four different cylinder radius distributions, indicating that the sub-voxel size should be chosen based on the dimension of the cylinders rather than arbitrarily.

\Fig{subvoxgridsize:v_40} shows the same plots, but for experiments performed with a packing density of $V_I\approx0.40$. In this case, the cylinder placement time is minimized for a sub-voxel size of 2$\bar{D}$ for all cylinder distributions. This is consistently larger than for $V_I\approx0.7$, due to the fact that each sub-voxel contains fewer cylinders and thus requires a smaller number of intersection checks when placing each cylinder. The time taken for the simulation dynamics is minimized when the sub-voxel size is $Mo(D)$ for the two smallest cylinder distributions and $Mo(\frac{D}{2})$ for the larger distributions. Once again, due to the dominance of the dynamics over the placement, the overall simulation run time is minimized for the same sub-voxel size that minimizes the simulation dynamics run time. 
As for the high volume fraction experiments, the minimum is very flat from $\frac{\bar{D}}{4}$ to $2\bar{D}$. Once again the curves for simulations run with different cylinder radius distributions follow the same trends, and the trends are consistent with those observed for $V_I\approx0.7$.

Overall, these results indicate that the run time is minimized when the size of the sub-voxels is between the mode cylinder radius and diameter, regardless of the cylinder radius distribution or packing density used here. The minimum is fairly broad and thus is not sensitive to small changes in sub-voxel size. We therefore use mode cylinder radius in the remainder of the experiments, as these are performed using the highest packing densities possible. We note that these results are specific to the choice of cylinder distribution and may not be valid for other, very different cylinder distributions such as highly bimodal distributions. For these types of distributions, this experiment should be repeated to determine the optimal sub-voxel size.

\subsection*{Sample size effects}
Larger substrates allow a larger number of samples from the model radius distribution. We would therefore expect sample bias and variation between cylinder radius histograms for different statistically-similar substrates to be reduced. To illustrate this effect we perform an experiment in sampling from a gamma distribution and then performing a maximum likelihood estimation of the distribution parameters. Since we know the ground-truth parameters for the underlying distribution, this allows us to measure accuracy and precision of parameter estimates as a function of sample size.

We draw 1000 sets of $M$ samples from $\Gamma(k,\theta)$ with $k=5.92$ and $\theta=1.06\times10^{-7}$m. We then perform a maximum likelihood estimation of the parameters $k$ and $\theta$ for each sample by using the method of Choi \& Wette \cite{ChoiWett1969}, which we briefly reproduce here.

The maximum likelihood estimation of $\theta$ is given by
\beq
	\hat{\theta}=\frac{1}{kM}\sum_{i=1}^{M}x_i.
\eeq
This expression for $\hat{\theta}$ can be substituted into the log-likelihood function, yielding an expression for $\hat{k}$. There is no simple closed form for the maximum likelihood estimation of $k$, however an approximate form is given by 
\beq
	\hat{k}\approx \frac{3-s+\sqrt{(s-3)^2+24s}}{12s},
\eeq
which is accurate to within 1.5\%, and may be refined by performing Newton-Raphson updates according to 
\beq
	\hat{k} \gets \hat{k}- \frac{\ln(\hat{k})-\psi(\hat{k})-s}{\frac{1}{\hat{k}}-\psi'(\hat{k})}
\eeq
where
\beq
	s=\ln\left(\frac{1}{M}\sum_{i=1}^Mx_i\right) - \frac{1}{M}\sum_{i=1}^M\ln(x_i)
\eeq
and $\psi(z)$ is the digamma function.

We estimate the maximum likelihood distribution parameters $\hat{k}$ and $\hat{\theta}$ for samples with $M$ ranging from $10^2$ to $10^6$. This range is chosen such that the lower end is representative of the sample sizes of previous simulation work, whereas the upper end is representative of the approximate number of white matter fibers in a typical scan voxel. We calculate the mean and standard deviation of parameter estimates for 1000 sets of samples at each size. 

\Fig{MLE} shows the mean parameter estimates and their standard deviations. As expected, the precision and accuracy of the estimates improves with increasing sample size. In particular, the standard deviations in both cases decrease by two orders of magnitude over the range considered. This illustrates that parameter estimates from smaller samples can be highly sample-specific and that larger sample sizes are highly desirable.

We observe a bias in parameter estimate means. For the smallest samples this bias is on the order of 3\%, falling to less than 0.1\% for sample sizes of 1000 or greater. The sample size required to reduce bias to a desired level is a function of the shape parameter of the gamma distribution (the scale parameter is a simple multiplying factor). Changing the shape parameter changes the relative weight of the tail of the distribution and therefore the number of samples required to probe it adequately. This is observed across a wide range of gamma distribution parameters - a lower value for the shape parameter generally acts to reduce the number of samples required to achieve a desired level of bias. The parameters shown here are representative of those observed in healthy human white matter tissue \cite{aboitiz}.

\subsection*{Variation in substrates}
\label{var:vi}
In this section, we test the effect of sample size on achievable packing density. We maximize packing density for a particular collection of $M$ cylinders by incrementally reducing the substrate size until algorithm \ref{alg:substrate} fails to pack all cylinders in. That substrate size defines the maximum achievable packing density. Substrate sizes for each $M$ are given in table-\ref{subsizes}. We repeat this process 30 times with different random draws of cylinders from a gamma distribution with parameters $k=5.92$ and $\theta=1.06\times 10^{-7}$m for $M=10^2,\ldots,10^6$.

\Fig{volfracs} shows the mean volume fraction for each substrate size, with error bars showing the range of values. For substrates with 1000 cylinders or greater, we observe a consistent mean volume fraction of around 0.7. The variation around the mean decreases with increasing $M$. 

The situation for $M=10^2$, however, is quite different. As shown in \fig{MLE}, small sample sizes lead to large variability the packing fractions achieved for different draws of radii. This illustrates that smaller sample sizes and the variability in maximum likelihood parameter estimates observed in the previous section can lead to broad variation in packing fractions achieved on the substrate. Smaller sample sizes can be less representative of the underlying desired radius distribution than larger ones, leading to sample bias and increased (or decreased) difficulty of the associated packing problem. It is therefore crucial that the number of cylinders on the substrate is large enough to avoid introducing simulation-based bias.

\subsection*{Calibrating simulation parameters}
\label{calibrate:simprams}
We now investigate the effect of substrate size on simulation parameters. Run-time is proportional to the number of spins $N$ and updates $T$. We therefore define a simulation complexity parameter $X$ as
\beq
  X=NT.
\eeq
Here we test whether the operating point of the trade-off between $N$ and $T$ for fixed $X$ depends on the substrate size.

We choose $X=5\times 10^8$ spin-updates, and run simulations with $N=10^1, 10^2, 10^3, 10^4,$ and $10^5$ with $T=\frac{X}{N}$ in each case. Substrates are constructed using algorithm 1 using the parameters from the Choice of Subvoxel Grid Size section. We run 30 repeats of each combination of $N$ and $T$ using different random seeds for both substrate construction and spin dynamics and repeat this for substrates with different numbers of cylinders $M$ where $M= 10^2, 10^3, 10^4, 10^5,$ and $10^6$ with substrate sizes given in table-\ref{subsizes}. Synthetic diffusion-weighted measurements are generated using the same procedure as \cite{hallAlexander09} using a PGSE sequence with parameters $\Delta=40$ms, $\delta=2$ms and $|\mathbf{G}|=112$mTm$^{-1}$ with gradients applied perpendicular to cylinder axes. Scan parameters chosen here are an arbitrary selection from the parameters ranges considered in the final experiment in this section. Results shown are representative of parameter choices across this range.

\begin{figure}[h!]
	\begin{center}
		\subfigure[Cylinder volume fractions as a function of substrate size]{
			\includegraphics[width=0.65\textwidth]{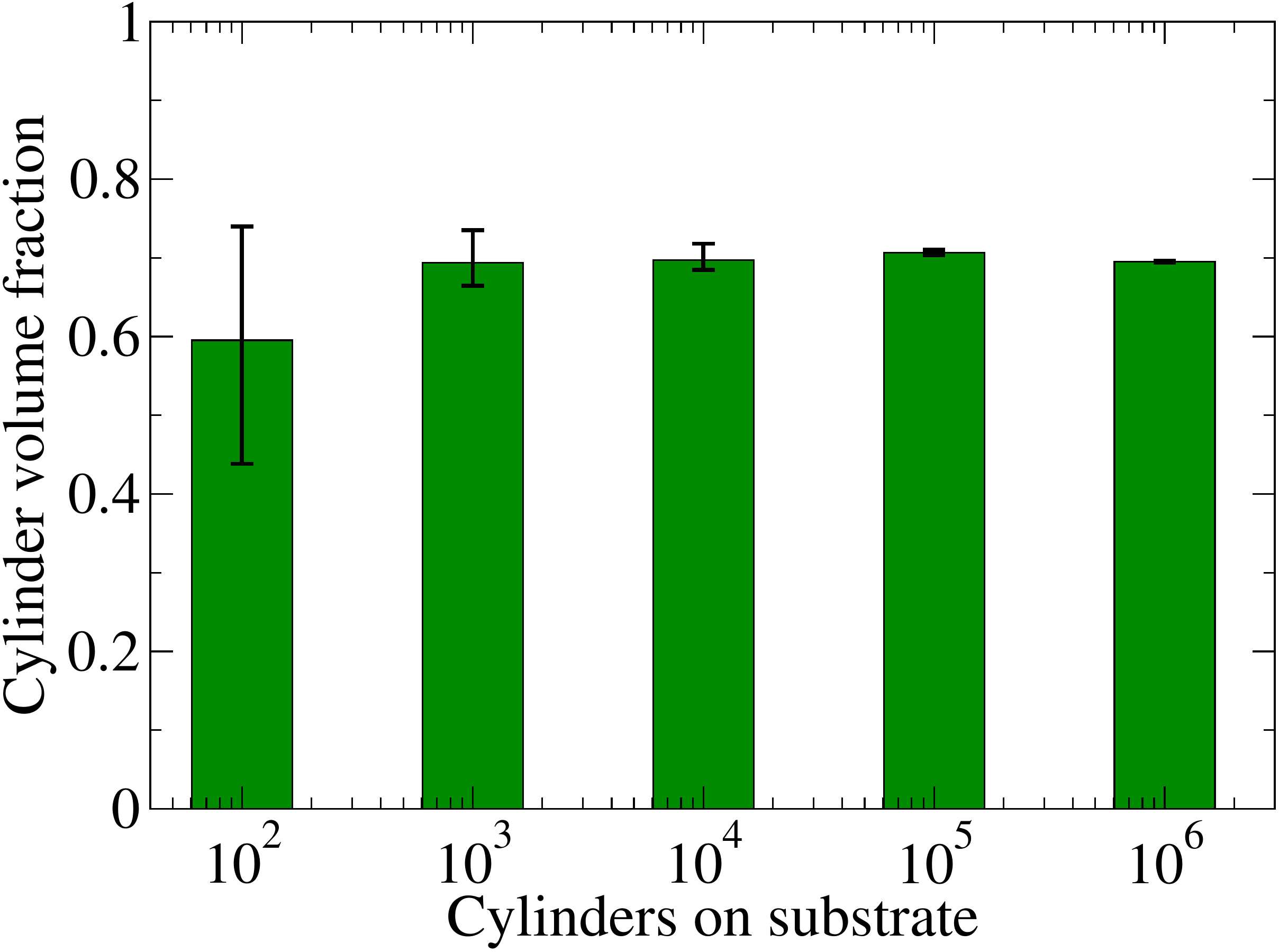}
			\label{volfracs}  
		}
	  
		\subfigure[Mean and standard deviation of diffusion-weighted signals as a function of number of spins]{
			\includegraphics[width=0.65\textwidth]{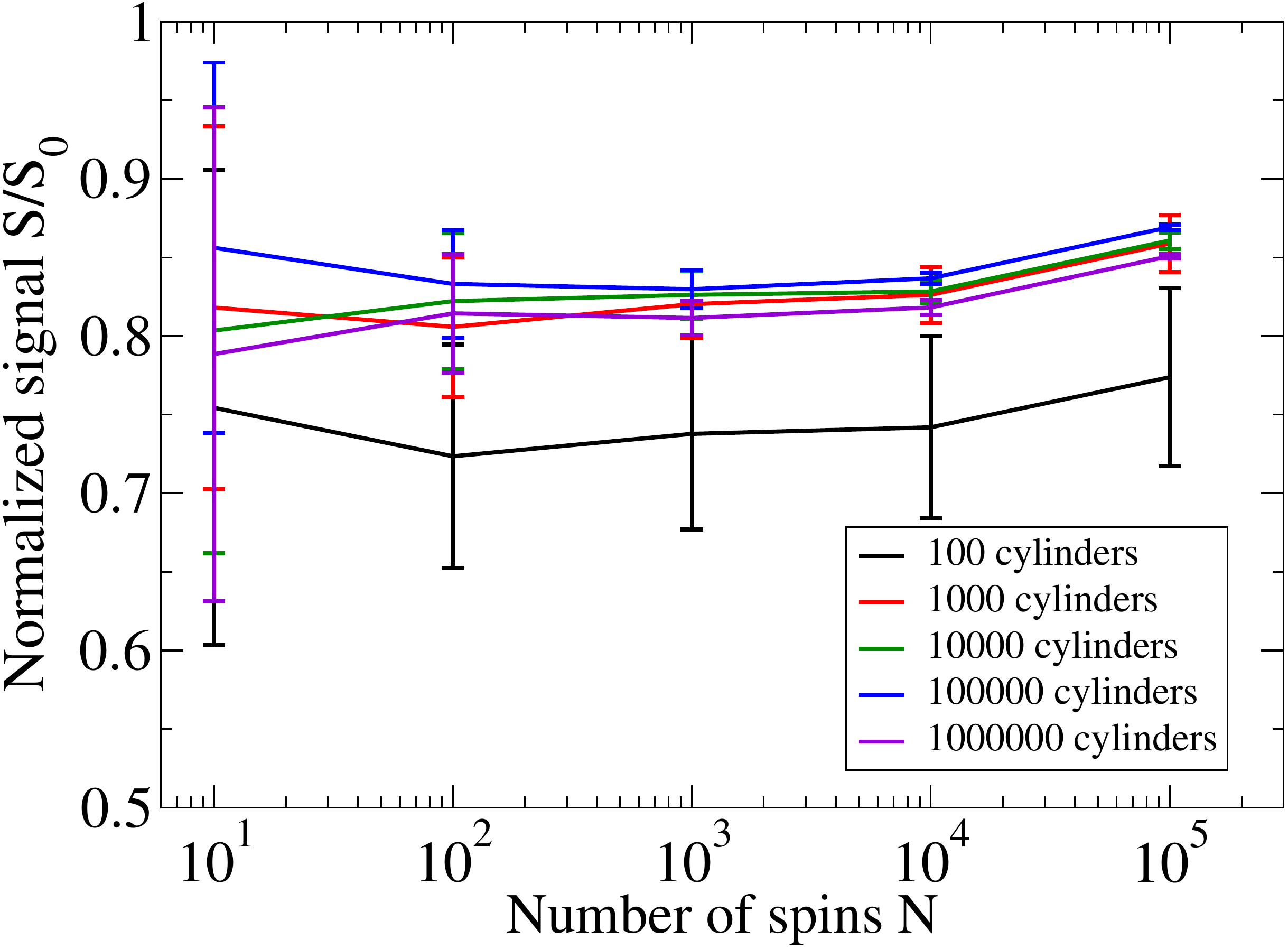}
			\label{hockeystick}
		}
	\end{center}
	\caption{Maximum cylinder volume fractions achieved as a function of the number of cylinders on the substrate and variation in diffusion-weighted signals as a function of number of spins in the simulation.}
	
\end{figure}

\Fig{hockeystick} shows the mean and standard deviation of synthetic signals as a function of number of spins for different numbers of cylinders. As observed by \cite{hallAlexander09} the variation in signals decreases with increasing number of spins $N$. This is observed for all substrate sizes considered. The most extreme variations are observed when $N=10$, as small $N$ leads to fluctuations due to poor sampling.

Again in common with \cite{hallAlexander09} we observe that the mean signal reduces with $N$ to a minimum before increasing again as a bias is introduced for large $N$. This is caused by large step lengths (due to small $T$) introducing an additional apparent restriction\cite{hallAlexander09}. The position of this minimum is consistent across all substrate sizes, occurring for $N=10^3$ and $N=10^4$. There is no significant difference between the mean signals at these points, but the standard deviations are smaller at $N=10^4$. We will therefore use $N=10^4$, $T=5\times 10^4$ in all further experiments.

For substrates containing 1000 cylinders or more there is little difference between the pattern of signals generated by simulations - error bars typically overlap and the overall trends (aside from at $N=10$) show similar overall patterns. In common with the cylinder volume fraction results, however, this is not the case when considering the smallest substrates. Here we observe consistently higher standard deviations and a consistent bias in mean. The consistency between \fig{hockeystick} and \fig{MLE} suggest that signal bias is the result of changes in cylinder volume fraction, and that larger substrates are required in order to ensure signal reproducibility from simulations.

\subsection*{Substrate size and spin dynamics}
\label{var:msd}
This section investigates how changes in the substrate due to sample and substrate size affect the intrinsic dynamics of diffusing spins. We investigate whether sample bias and substrate size have a direct effect on diffusion dynamics on the substrate. 

We consider statistical measures of spin displacement over 30 different realizations of substrates of each size. We refer to each set of realizations as an ``ensemble''. The quantities of interest are the ensemble average of the mean squared-displacement of spins and the ensemble standard deviation of the mean-squared displacement, both as functions of time.

\begin{figure}[h!]
  \begin{center}
    \subfigure[Ensemble average mean squared displacements]{
      \label{msd:msd}
      \includegraphics[width=0.45\textwidth]{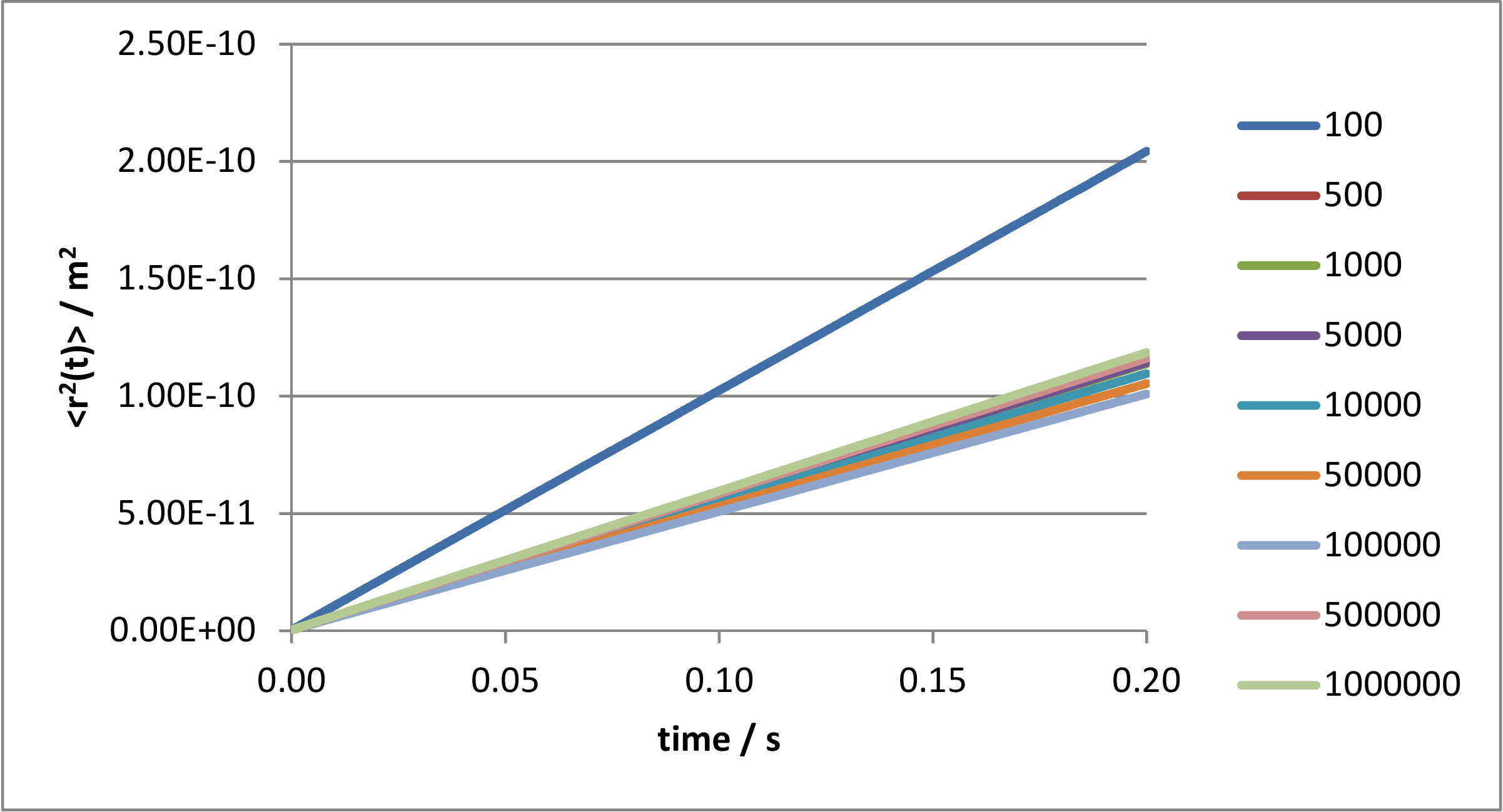}
    }
	\\
    \subfigure[Time-normalized ensemble average mean squared displacements]{
      \label{msd:diff}
      \includegraphics[width=0.45\textwidth]{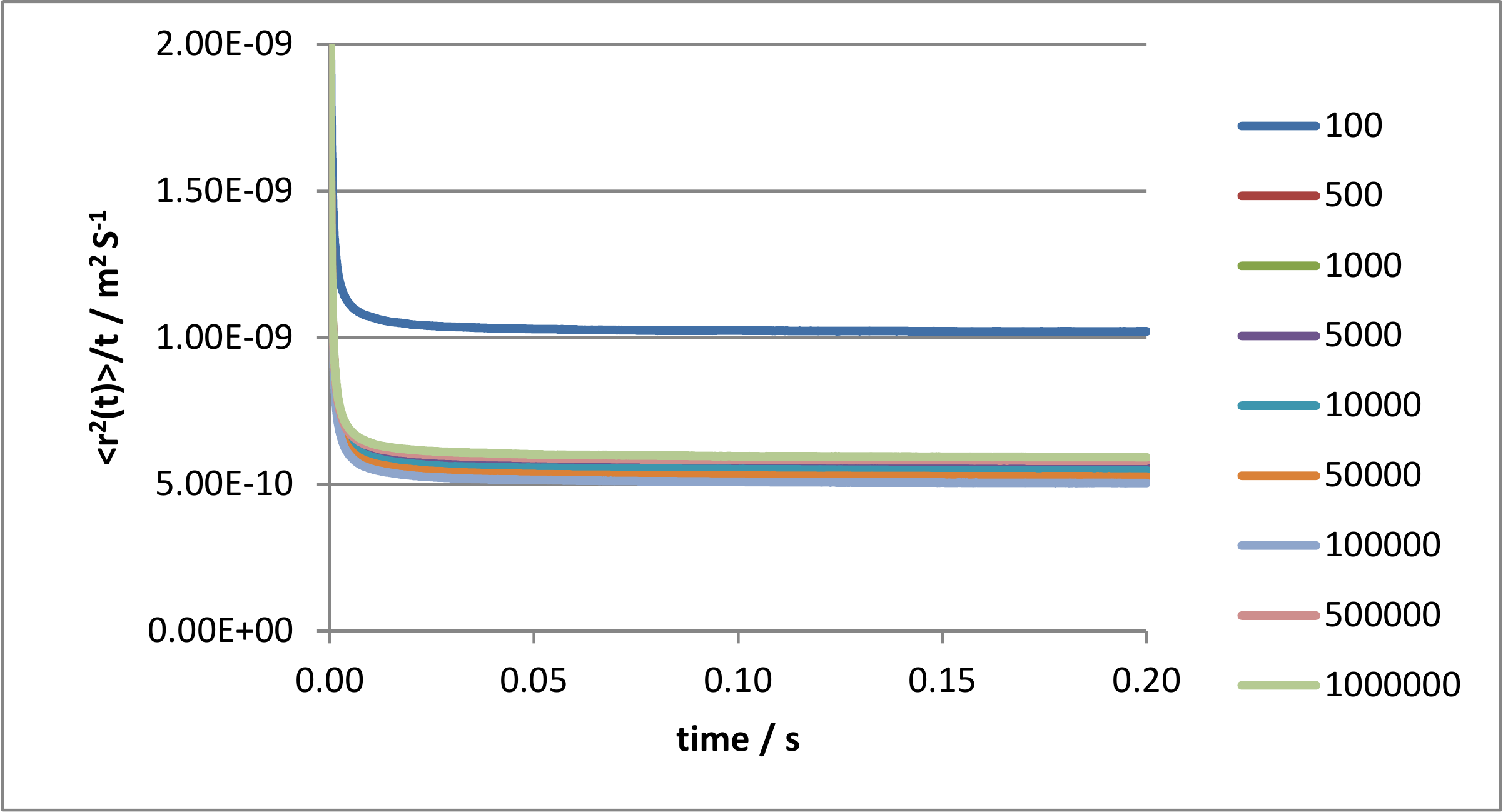}
    }
	\\
    \subfigure[Time-normalized ensemble standard deviation mean squared displacements]{
      \label{msd:sig}
      \includegraphics[width=0.45\textwidth]{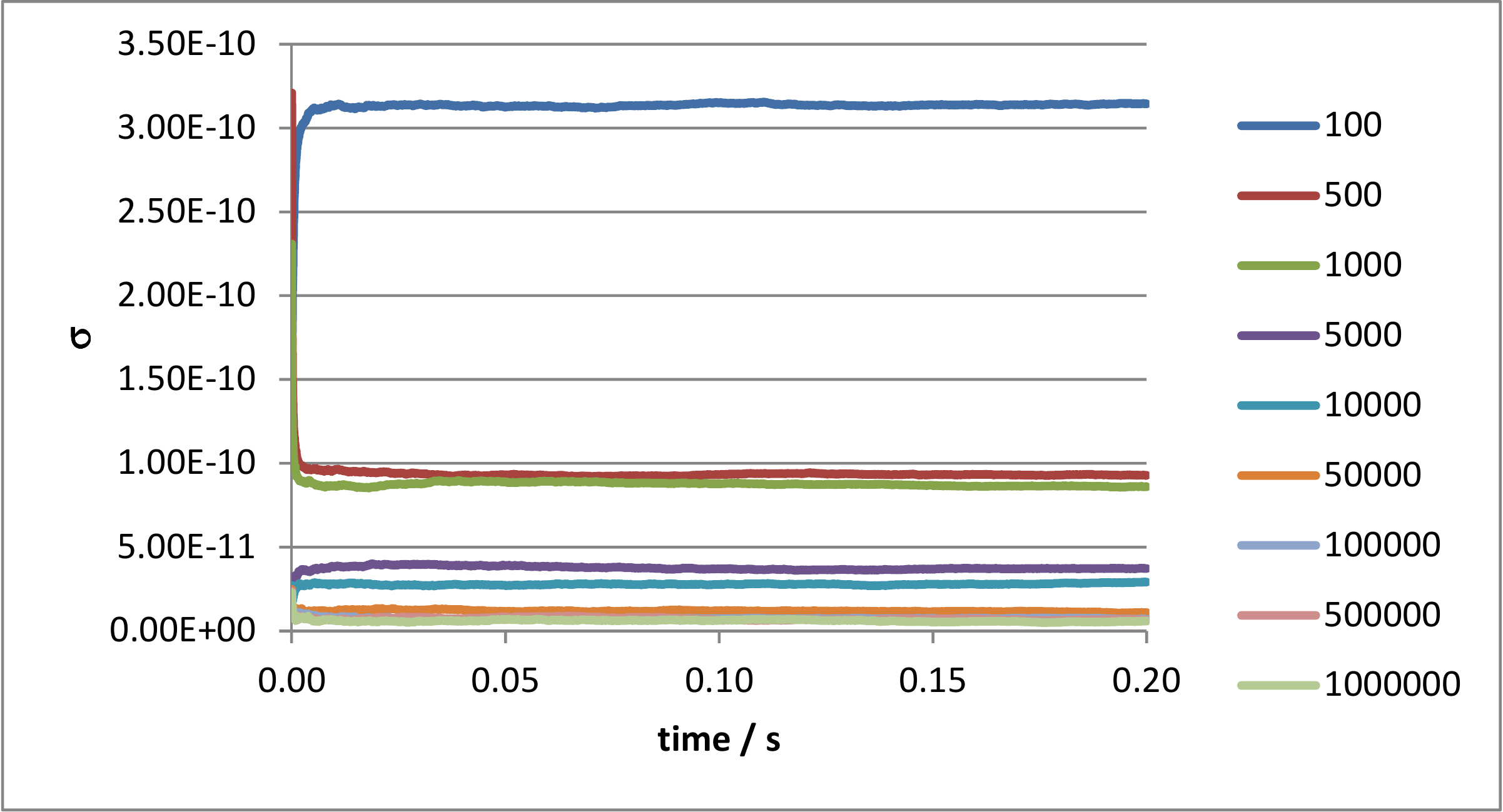}
    }
    \caption{The effect of substrate size on spin diffusion dynamics. The smallest substrate size shows differences in its dynamics compared to diffusion on other substrates.}
	\label{msd}

  \end{center}
\end{figure}

\Fig{msd}(a) shows the ensemble average mean-squared displacement of spins as a function of time for substrates with $M=10^2,...,10^6$. With the exception of the substrate with $M=10^2$, ensemble averaged mean squared-displacements exhibit similar behaviors and similar slopes. Mean squared-displacements for $M=10^2$ deviate considerably from the remainder, with spins exhibiting larger mean squared displacements and a steeper slope. This difference is more apparent in (b), which shows the ensemble average mean squared displacements normalized by time. This gives a measure of spin diffusivity at a particular time. We observe the same overall profile in all curves: the short time limit of free diffusion rapidly decreasing and plateauing to a long-time limiting value over a period of approximately 5ms. The diffusivity of spins in the substrate with $M=10^2$ is higher than that observed for all other curves, which is to be expected given the differences in cylinder volume fraction.

\Fig{msd}(c) shows the time-normalized ensemble standard deviation of the mean-squared displacements. We observe a decrease in the asymtotic value with increasing $M$, illustrating that variation in diffusion dynamics between different samples of the same size is reduced with increasing the number of cylinders.

\subsection*{Variation in synthetic diffusion-weighted data}
\label{var:dwi}
We now investigate the variation in diffusion-weighted signals synthesized from Monte-Carlo simulations with realistic voxel sizes. Substrates are constructed using the same parameters used in the Choice of Subvoxel Grid Size section. 

Synthetic diffusion-weighted signals are calculated as in \cite{hallAlexander09}: diffusion-weighted signals are a function of all spin trajectories in the simulation. We consider only the smallest and largest substrates from the previous experiment ($M=10^2$, $L=1.55\times 10^{-5}$ m and $M=10^6$, $L=1.56\times 10^{-3}$ m).

We generate synthetic measurements across a range of parameters. Measurements are generated for gradient strength $\mathbf{G}\in\{0.01, 0.02, \ldots, 1\}$Tm$^{-1}$, diffusion time $\Delta\in\{0.01, 0.02, \ldots, 0.1\}$s, and gradient pulse duration $\delta= 0.002$s for all combinations such that $\delta\le\Delta$. We repeat the simulations for each substrate 30 times using different random seeds and obtain the mean and standard deviation of measurements at each set of scan parameters. 

\begin{figure}[h!]
  \begin{center}
    \includegraphics[width=0.55\textwidth]{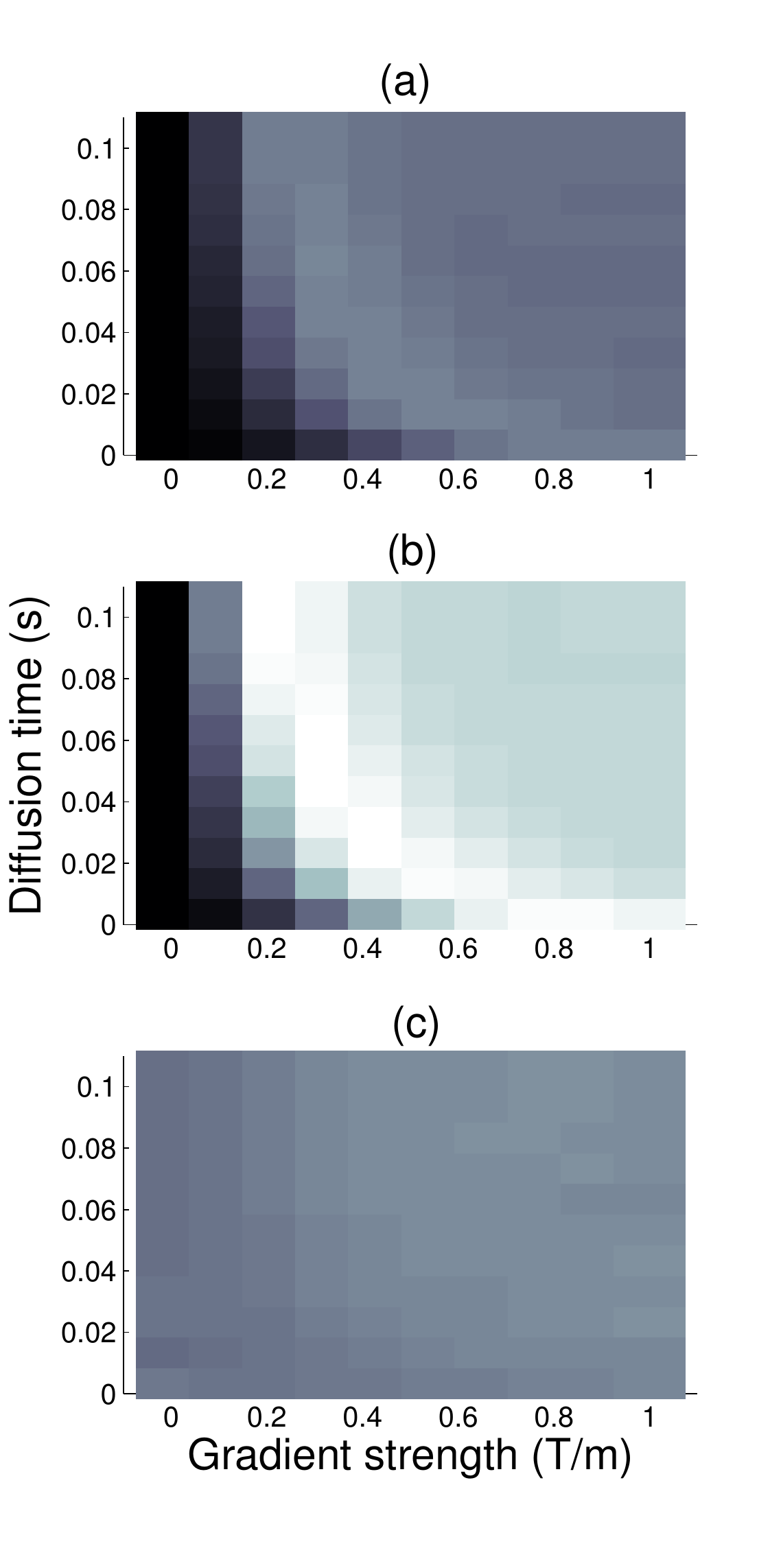}
    \caption{The standard deviation of synthesized diffusion-weighted measurements as a function of gradient strength and diffusion time for (a) 1,000,000 cylinders, and (b) 100 cylinders. (a) and (b) are shown in the same color-scale with black zero and white equal to the unweighted signal. (c) Shows the ratio between standard deviations.}
    \label{vardwi}
  \end{center}
\end{figure}

\Fig{vardwi} shows the standard deviations of diffusion-weighted signals generated perpendicular to the cylinder axis as a function of diffusion time and gradient strength. Higher intensity corresponds to greater variation. (a) shows results from the substrate with $M=10^6$, (b) shows results from the substrate with $M=10^2$; both are shown on the same color scale. (c) shows the ratio of (a) and (b) at each parameter combination. 

The overall pattern of variation is consistent across both substrates: at low diffusion-weightings the variation is low, the variation increases rapidly to a ridge which follows a line of constant $b$-value. At higher diffusion weightings the variation slowly decreases. Variation in the substrate with $M=10^6$ is consistently lower than the substrate with $M=10^2$. \Fig{vardwi}(c) shows the ratio of the two standard deviations as a function of the scan parameters. The ratio is an approximately constant value of 2.1, increasing by a small amount with increasing diffusion weighting. 

\section*{Discussion \& Conclusion}
\label{concs}
We have shown that optimizing intersection checking by using a spatial subdivision algorithm hugely improves the efficiency of Monte-Carlo simulation during substrate construction as well as dynamics simulation and shown how to choose a sub-voxel grid size to minimize run-time. The larger and more complex substrates that the technique makes possible reduce variation and bias in synthetic diffusion-weighted measurements. 

Using the spatial subdivision algorithm, the performance of the simulation is no longer dictated by the number of objects in the substrates but by the number of objects in a typical sub-voxel. The more objects in the sub-voxel, the more expensive the intersection check. Conversely, the more subvoxels used, the more memory required and the more subvoxels will be intersected by each object. This limits the performance gains available and ultimately causes additional overheads to avoid checking the same object multiple times. We have shown (see \fig{subvoxgridsize}) that this is optimized when the size of sub-voxels is between the mode of the cylinder radius and diameter; however the minimum is broad. 

\begin{figure}[h!]
  \begin{center}
    \subfigure[Estimates of $k$ as a function of sample size.]{
      \label{MLE_k}
      \includegraphics[width=0.85\textwidth]{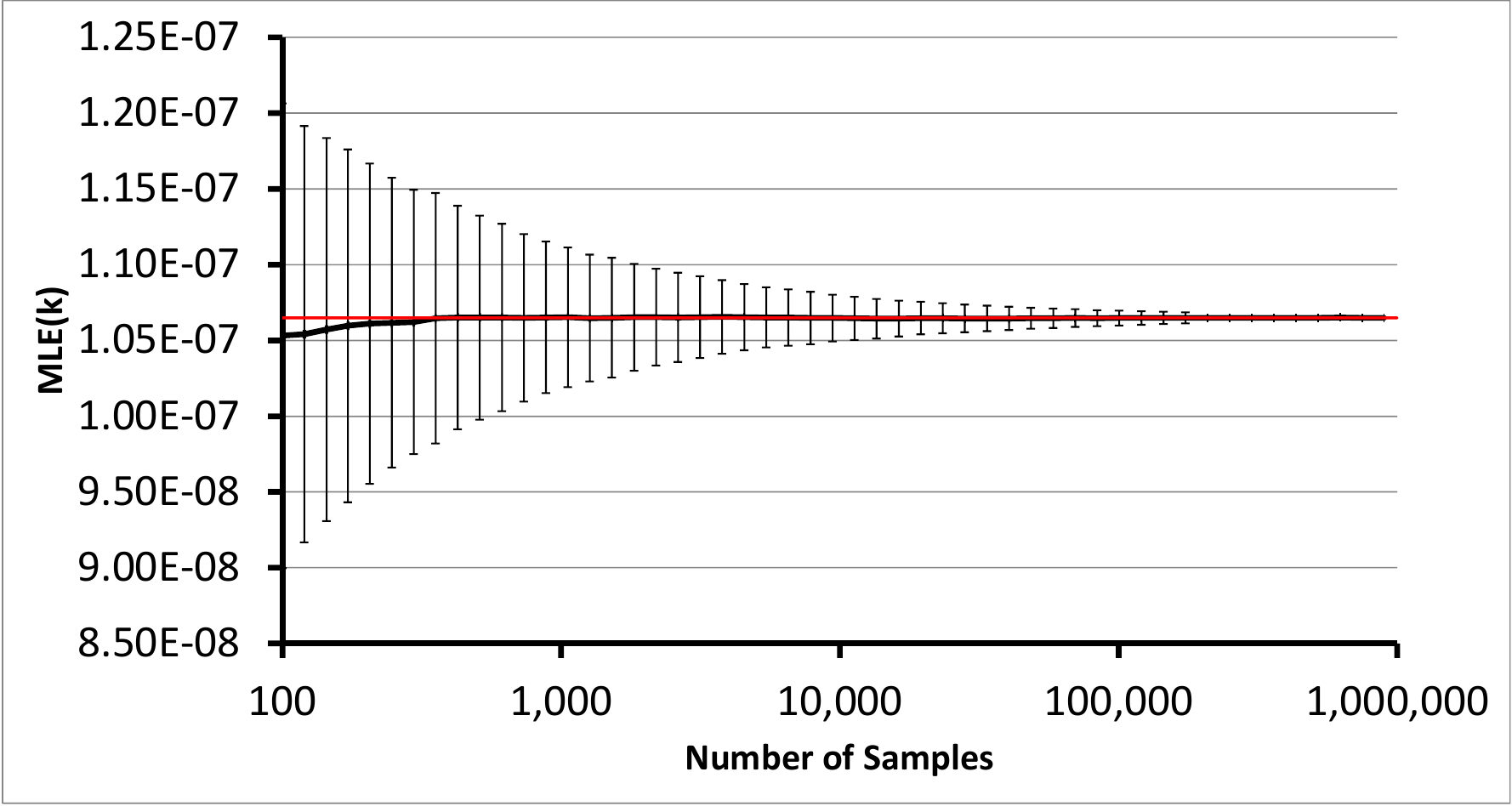}
    }
    \subfigure[Estimates of $\theta$ as a function of sample size.]{
      \label{MLE_beta}
      \includegraphics[width=0.85\textwidth]{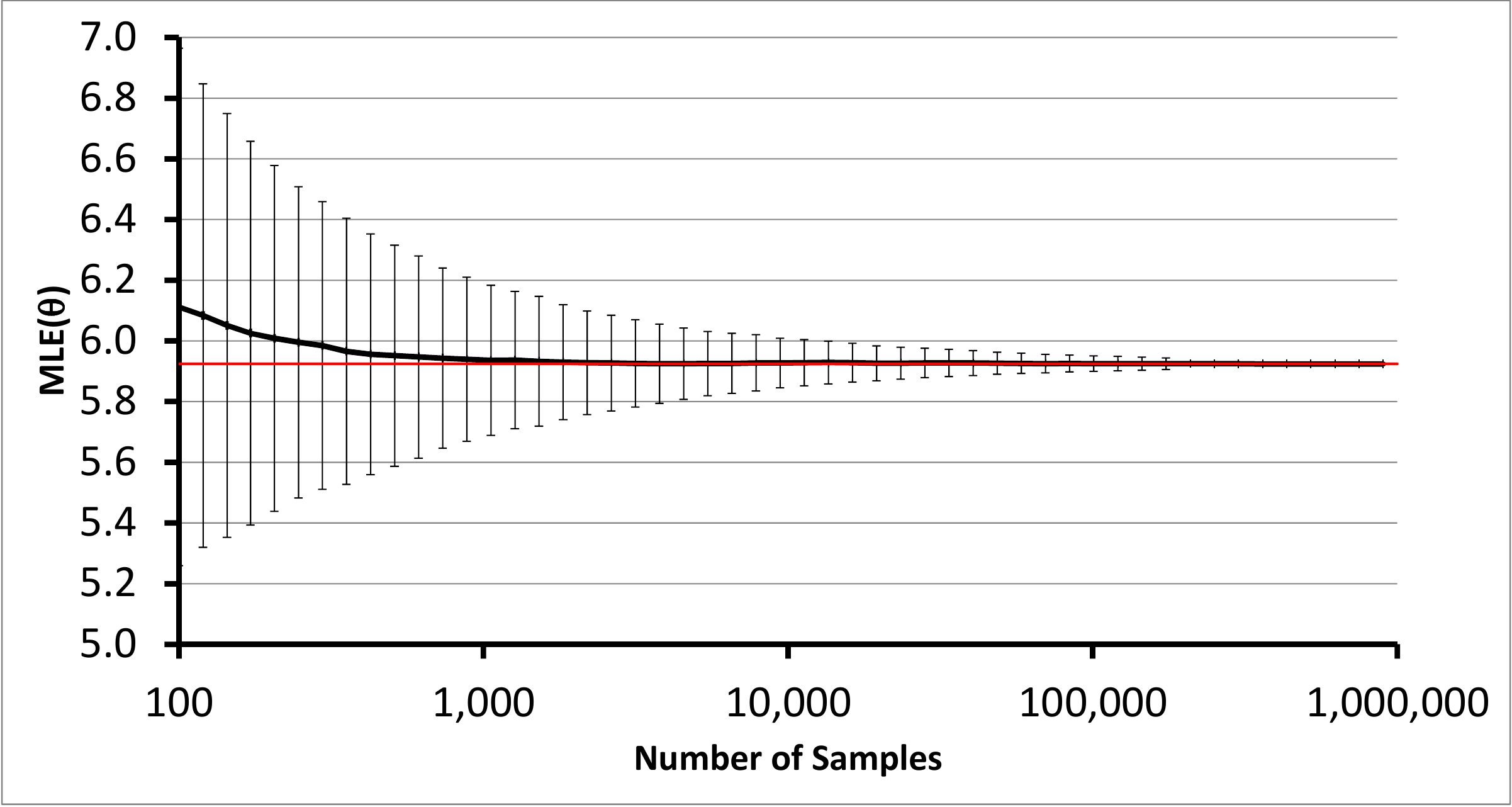}
    }
    \caption{The effect of sample size on the accuracy and precision of gamma distribution parameters. Increasing the sample size increases both accuracy and precision of maximum likelihood estimates of distribution parameters. Black line shows the mean parameter estimate, error bars show $\pm$ one standard deviation, ground truth parameter values shown in red.}
    \label{MLE}
  \end{center}
\end{figure}

As spatial subdivision decouples the run-time from the total number of objects, we can now run simulations using substrates that contain large samples of cylinders. This is important as sample size is a significant source of bias and variation. For example, with the gamma distributions we use here, \fig{MLE} shows the sample bias from $M=100$ is substantial, which is typical of sample sizes in previous simulation work. It also has a direct impact on the packing fraction achievable on a given substrate (\fig{volfracs}). Smaller sample sizes on smaller substrates lead to larger bias and variation in intracellular volume fractions compared to larger ones. The volume fraction has a direct effect on extracellular diffusion (\fig{msd}) and hence on the diffusion-weighted signal. Sample variation therefore leads to variation in signals synthesized from simulation. 

It is therefore imperative that sample bias be kept to a minimum when synthesizing diffusion-weighted data from simulation. For cylinder radius distribution parameters typical of healthy human white matter, the sample size required to reduce bias below 1\% may be as high as 1000 (\fig{MLE}). This is an order of magnitude greater than has typically been considered in previous simulation work; however spatial subdivision makes simulations with substrates of this size easily achievable. It therefore leads to a reduction in bias and variation in synthetic diffusion-weighted signals (\fig{vardwi}), allowing more reliable data to be synthesized. This means that substrate sizes comparable to the size of in vivo scan voxels are accessible to widely available, desktop PC-level hardware. 

Although we have demonstrated the technique using cylinder-based substrates, uniform spatial subdivision can be applied to substrates comprised of any geometrical primitive. Mesh-based substrates, for example, can be thought of as collections of triangles. The sub-voxel map discussed in the methods section can be used just as effectively with triangles as with cylinders. Choosing an optimal sub-voxel grid size would require an investigation similar to that presented here, but the current work suggests that a grid size similar to the mean triangle size may be appropriate.

The choice of a specifically uniform subdivision assumes that objects are approximately evenly distributed across the substrate, which suits the substrates considered here. Some substrates may not suit uniform subdivision - for example those with dense clusters of smaller objects with large spaces or sparse regions of larger objects in between them. Under these circumstances uniform subdivision may be sub-optimal and a more hierarchical scheme such as a BSP-tree \cite{slater} may be more efficient. Even here, however, uniform subdivision will still lead to improved performance over a brute-force approach.

The difference in run-time between simulations with and without spatial subdivision is difficult to quantify for practical reasons: simulations with $M=1,000,000$ cannot be run in a feasible amount of time using desktop hardware without spatial subdivision. The slowest feasible runs considered here have a subvoxel grid size of twenty times the mean cylinder diameter - these are an order of magnitude slower than the fastest runs (30,000 seconds vs. 3,000 seconds) but still make use of spatial subdivision. Unoptimized runs were attempted, but failed to complete substrate construction after several days and were deemed impractical.

Given the observed differences in run-time, it is difficult to overestimate the importance of efficient intersection checking on simulations of this type. The implementation used in this work is written in Java and is freely available for download as part of the Camino diffusion MRI toolkit (\url{http://www.camino.org.uk/}), fully integrated with the Camino Monte-Carlo simulation framework. The implementation uses mean cylinder radius as sub-voxel size by default.

\section*{Acknowledgments}
During this work, MGH and DCA were supported by EPSRC grant number EP/E007748, GNG was supported by EP/I027084/01. 


\begin{thebibliography}{10}

\bibitem{aboitiz}
Francisco Aboitiz, Arnold~B. Scheibel, Robin~S. Fisher, and Eran Zaidel.
\newblock Fiber composition of the human corpus callosum.
\newblock {\em Brain Research}, 598(1–2):143 -- 153, 1992.

\bibitem{activeAx}
Daniel~C. Alexander, Penny~L. Hubbard, Matt~G. Hall, Elizabeth~A. Moore,
  Maurice Ptito, Geoff J.~M. Parker, and Tim~B. Dyrby.
\newblock Orientationally invariant indices of axon diameter and density from
  diffusion mri.
\newblock {\em Neuroimage}, 52(4):1374--1389, 2010.

\bibitem{Amanatides1987}
John Amanatides and Andrew Woo.
\newblock A fast voxel traversal algorithm for ray tracing.
\newblock {\em In Eurographics ’87}, pages 3--10, 1987.

\bibitem{BallFranks}
Gregory~T. Balls and Lawrence~R. Frank.
\newblock A simulation environment for diffusion weighted mr experiments in
  complex media.
\newblock {\em Magnetic Resonance in Medicine}, 62(3):771--778, 2009.

\bibitem{Best1978}
D.J. Best.
\newblock Letter to the editor.
\newblock {\em Applied Statistics}, 27, 1978.

\bibitem{budde}
Matthew~D. Budde and Joseph~A. Frank.
\newblock Neurite beading is sufficient to decrease the apparent diffusion
  coefficient after ischemic stroke.
\newblock {\em Proceding of the National Academy of Sciences USA},
  107(32):14472--14477, 2010.

\bibitem{ChoiWett1969}
S.~C. Choi and R.~Wette.
\newblock {Maximum Likelihood Estimation of the Parameters of the Gamma
  Distribution and Their Bias}.
\newblock {\em Technometrics}, 11(4):683--690, 1969.

\bibitem{ClearyWyvill1988}
John~G. Cleary and Geoff Wyvill.
\newblock Analysis of an algorithm for fast ray tracing using uniform space
  subdivision.
\newblock {\em The Visual Computer}, 4(2):65--83, 1988.

\bibitem{devroye1986}
L.~Devroye.
\newblock {\em Non-uniform Random Variate Generation}.
\newblock Springer-Verlag, New York, 1986.

\bibitem{ElsSim}
Els Fieremans, Yves~De Deene, Steven Delputte, Mahir~S. Özdemir, Yves
  D’Asseler, Jelle Vlassenbroeck, Karel Deblaere, Eric Achten, and Ignace
  Lemahieu.
\newblock Simulation and experimental verification of the diffusion in an
  anisotropic fiber phantom.
\newblock {\em Journal of Magnetic Resonance}, 190(2):189 -- 199, 2008.

\bibitem{fujimoto1986}
A.~Fujimoto, Takayuki Tanaka, and K.~Iwata.
\newblock Arts: Accelerated ray-tracing system.
\newblock {\em Computer Graphics and Applications, IEEE}, 6:16--26, 1986.

\bibitem{hallAlexander09}
Matt~G. Hall and Daniel~C. Alexander.
\newblock Convergence and parameter choice for monte-carlo simulations of
  diffusion mri.
\newblock {\em Medical Imaging, IEEE Transactions on}, 28(9):1354--1364, 2009.

\bibitem{nilsson}
Markus Nilsson, Jimmy L{\"a}tt, Freddy St{\aa}hlberg, Danielle Westen, and
  H{\aa}kan Hagsl{\"a}tt.
\newblock The importance of axonal undulation in diffusion mr measurements: a
  monte carlo simulation study.
\newblock {\em NMR in Biomedicine}, 25(5):795--805, 2012.

\bibitem{NovikovNatPhys}
Dmitry~S. Novikov, Els Fieremans, Jens~H. Jensen, and Joseph~A. Helpern.
\newblock Random walks with barriers.
\newblock {\em Nature Physics}, 7:508--514, 2011.

\bibitem{LauraMICCAI2010}
Eleftheria Panagiotaki, Matt~G. Hall, Hui Zhang, Bernard Siow, Mark~F. Lythgoe,
  and Daniel~C. Alexander.
\newblock High-fidelity meshes from tissue samples for diffusion mri
  simulations.
\newblock {\em Medical Image Computing and Computer-Assisted
  Intervention--MICCAI 2010}, pages 404--411, 2010.

\bibitem{alonso}
Alonso Ramirez-Manzanares, Philip~A. Cook, Matt~G. Hall, Manzar Ashtari, and
  James~C. Gee.
\newblock Resolving axon fiber crossings at clinical b-values: An evaluation
  study.
\newblock {\em Medical physics}, 38(9):5239--5253, 2011.

\bibitem{slater}
M.~Slater, A.~Steed, and Y.~Chrysanthou.
\newblock {\em Computer Graphics and Virtual Environments: From Realism to
  Real-time}.
\newblock Addison-Wesley, 2002.

\end{thebibliography}

\pagebreak
\section*{Tables}

\begin{table}[!h]
  \begin{center}
	  \caption{Numbers of cylinders and substrate sizes used to assess the impact of substrate size on spin dynamics}
	  \begin{tabular}{c|c}
		Number of cylinders & Substrate size \\
		\hline
		1,000,000 & $1.46\times 10^{-3}$m\\
		500,000 & $1.03\times 10^{-3}$m\\
		100,000 & $4.55\times 10^{-4}$m\\
		50,000 & $3.23\times 10^{-4}$m\\
		10,000 & $1.45\times 10^{-4}$m\\
		5,000 & $1.03\times 10^{-4}$m\\
		1,000 & $4.57\times 10^{-5}$m\\
		500 & $3.20\times 10^{-5}$m\\
		100 & $1.55\times 10^{-5}$m\\
	  \end{tabular}
    \label{subsizes}
 \end{center}
\end{table}

\pagebreak
\section*{Figures}

\end{document}